\documentclass{article}

    \PassOptionsToPackage{numbers, compress}{natbib}

\usepackage[preprint]{neurips_2023}

\usepackage[]{mdframed}

\usepackage[utf8]{inputenc} 
\usepackage[T1]{fontenc}    
\usepackage{hyperref}       
\usepackage{url}            
\usepackage{booktabs}       
\usepackage{amsfonts}       
\usepackage{nicefrac}       
\usepackage{microtype}      
\usepackage{xcolor}         

\usepackage{amsmath,amsfonts,amsthm}
\usepackage{amsmath,bm}
\usepackage{amssymb}
\usepackage{subfigure}
\usepackage{graphicx}
\usepackage{enumitem} 
\usepackage{algorithmic,algorithm}
\usepackage{multirow}

\usepackage[utf8]{inputenc} 
\usepackage[T1]{fontenc}    
\usepackage{hyperref}       
\usepackage{url}            
\usepackage{booktabs}       
\usepackage{amsfonts}       
\usepackage{nicefrac}       
\usepackage{microtype}      
\usepackage{xcolor}         
\usepackage{enumitem}

\title{GPT-4 Vision on Medical Image Classification -- A Case Study on COVID-19 Dataset}

\author{%
  Ruibo Chen,~
  Tianyi Xiong,~
  Yihan Wu,~
  Guodong Liu,~
  Zhengmian Hu,~\AND
   Lichang Chen,~
  Yanshuo Chen,~
  Chenxi Liu,~
  Heng Huang\AND
  \\University of Maryland
}

\begin{document}

\maketitle

\section{Introduction}

In the intricate landscape of modern healthcare, medical image classification emerges as a pivotal task, driving crucial decisions in diagnosis, treatment planning, and patient management. This process involves the systematic categorization of various types of medical imagery—including X-rays, CT scans, MRIs, and ultrasound—into distinct classes that assist healthcare professionals in identifying anomalies, understanding physiological phenomena, and detecting diseases at early stages. The reliability and precision of image classification are paramount, given that these determinations form the bedrock upon which medical practitioners build their diagnostic and therapeutic strategies, directly impacting patient outcomes. With an increasing influx of complex imaging data and a growing need for rapid, accurate interpretation, the medical sector faces significant pressure to evolve beyond traditional analysis methods, necessitating innovative solutions that enhance the efficiency and accuracy of image classification.

The advent of large foundation models in artificial intelligence has ushered in a transformative era of computational capabilities. These models, characterized by their extensive scale, diverse training datasets, and impressive adaptability, have demonstrated profound impacts across various domains. Within the realm of medical image classification, there is burgeoning curiosity around the potential applicability and benefits of these formidable tools. The traditional approach, reliant on Convolutional Neural Networks (CNNs) based architectures, such as VGG \cite{simonyan2014very}, inception \cite{szegedy2015going}, ResNet \cite{he2016deep}, and DenseNet \cite{huang2017densely} have achieved noteworthy success in image categorization tasks \cite{cai2020review}. However, these methods often require vast amounts of labeled data and substantial computational resources, besides lacking the intuitive adaptability inherent in human cognition. Besides training a neural network from end to end, some transfer learning and self-supervised learning techniques are also employed in the filed of medical image classification to improve the efficiency and performance \cite{kim2022transfer,huang2023self}. But they are also limited by the predictive capability and few- or zero-shot learning ability \cite{nayem2023few}. Recently, large foundation models, with their sophisticated understanding of nuanced patterns, offer a promising alternative, hypothesized to enhance the precision and context-awareness in classifying medical images, provided they can be effectively adapted to understand and interpret complex visual medical data.

This study ventures into the novel application of in-context learning strategies with GPT-4V, a derivative of the generative pre-trained transformer models, specifically oriented towards visual tasks. In-context learning allows the model to utilize prompts—minimal yet specific pieces of information or instructions—to guide its responses in performing a particular task, relying on its vast pre-trained knowledge base rather than traditional task-specific training. By harnessing this approach, we aim to tailor GPT-4V’s capabilities to interpret and classify medical images, an endeavor scarcely explored in existing literature. Our methodology involves the meticulous design of context-rich prompts that facilitate the model's understanding of medical imaging classifications. The preliminary results are striking, showing that our adapted GPT-4V model, when equipped with well-crafted prompts, can achieve classification accuracy comparable to established baseline models. This finding not only underscores the versatility of large foundation models in medical applications but also heralds a potentially more resource-efficient and adaptable future for medical image analysis.

\section{Related Work}

\textbf{In-context Learning:}
In-context learning~(ICL)~\cite{liu-etal-2022-makes, wei2021finetuned, wang2023large, chen2023demonstrations, role2022} is a paradigm that has recently gained prominence, particularly in the realm of LLMs~\cite{brown2020language, chowdhery2022palm, thoppilan2022lamda, rae2021scaling, radford2021learning}. This approach provides an efficient way for pre-trained models to understand and execute a task using demonstrations, e.g., input-output pairs, without resorting to extensive fine-tuning or retraining on task-specific data. 



The effectiveness of ICL is rooted in a phenomenon known as "few-shot learning." In traditional machine learning, models often require substantial amounts of labeled data specific to each task \cite{radford2021learning}. However, in few-shot scenarios~\cite{finn2017model}, a model uses only a minimal number of examples to understand a task. This process is akin to the way humans often learn—by relating new information to existing knowledge. In-context learning takes this a step further, often operating in "zero-shot" or "one-shot" contexts, where the model is either not provided with any task-specific examples or just a single one, respectively.

In medical image classification, the application of ICL is still nascent. The potential of large language models like GPT-3 or GPT-4 to generalize learning from textual contexts to more complex multi-modal tasks, including image classification, offers promising avenues for exploration. The key lies in the careful design of prompts that succinctly yet comprehensively convey the task rules and criteria, enabling the model to apply its pre-existing knowledge effectively across domains.

\textbf{Medical Image Classification:}
Historically, before the rise of deep learning methodologies, the domain of medical image analysis was dominated by manual processes and intricate feature extraction techniques. These handcrafted features, exemplified by histograms of oriented gradients (HOG) \cite{dalal2005histograms} and local binary patterns (LBP) \cite{he1990texture}, offered a structured but somewhat constrained approach to image interpretation. However, the deep learning revolution that began around 2006 introduced a transformative era, significantly altering the landscape of medical image classification. Modern models, trained on vast datasets, are adept at recognizing patterns and anomalies within medical imagery, such as X-rays, MRIs, and CT scans. The primary emphasis has shifted towards enhancing accuracy, improving interpretability, and sharpening the ability to detect subtle indicators of medical conditions. This facilitates healthcare professionals in making more informed decisions. For instance, in the classification of retinal fundus images \cite{ganjdanesh2022multi,TAPE}, the use of strategies like data augmentation and transfer learning, especially with renowned pre-trained models like VGG19, has notably improved the detection accuracy of fundus diseases. Another pivotal task has been the detection of COVID-19 from chest X-rays. Recent advancements, which leverage the feature fusion of DenseNet and VGG, have demonstrated superior performance in detecting COVID-19 patients \cite{kong2022classification}.

\section{Our Method}

Our method's intuition lies in the meticulous crafting of prompts that serve as input for the advanced GPT-4V model, guiding it to make accurate medical image classifications. By employing in-context learning, we harness the model's expansive pre-existing knowledge base, prompting it with both text and images for robust task performance.

The experiment contrasts two central methodologies. The baseline involves traditional medical image analysis using ResNet models. Our approaches start with the naive zero-shot prompt on GPT4V, then enhance the naive prompt with more sophisticated approaches, meticulously structured to "complete the story" that the image tells.

In our experiments, we primarily utilized three categories of prompts to engage GPT-4V in the classification tasks: 
\begin{enumerate}
    \item A straightforward method, wherein GPT-4V is directly instructed to classify the presented images, representing the naive approach.
    \item An in-context learning strategy, which involves providing GPT-4V with multiple labeled examples to facilitate guided learning.
    \item A more nuanced in-context learning process that incorporates reasoning, requiring us to elucidate the relationships between the images and their corresponding labels for GPT-4V.
\end{enumerate}

The specific prompts employed for these methodologies are detailed subsequently:
\begin{mdframed}
\noindent\textbf{Naive approach.} Zero shot inference on GPT4V:\\
\noindent Upload an image on GPT4V.\\
\noindent \textbf{Prompt:} ``Instruction: classify the image into two classes {class1, class2}.
Please first output one line for the label of the image. In the subsequent line, please provide a comprehensive explanation of your classification." 
\end{mdframed}

\begin{mdframed}
\noindent\textbf{In-context learning 1 (ICL1).}\\
\noindent Upload three images separately on GPT4V, two for in-context learning and one for predicting.\\
\noindent \textbf{Prompt:} ``Instruction: classify the images into two classes {class1, class2}

\noindent Example: the label of the above images:\\
\noindent Image 1: class1\\
\noindent Image 2: class2\\
\noindent Please first output one line for the label of image 3. In the subsequent line, please provide a comprehensive explanation of your classification." 
\noindent Drawback: the attention of GPT4V may concentrate on certain image and lead to biased results. Thus, in the next ICL prompt we will try to put all images in one figure.
\end{mdframed}

\begin{mdframed}
\noindent\textbf{In-context learning 2 (ICL2).}\\
\noindent Combining three images on one figure and uploading it on GPT4V, two for in-context learning and one for predicting.\\
\noindent \textbf{Prompt:} ``Instruction: classify the images into two classes {class1, class2}

\noindent Example: the label of the above images:\\
\noindent Image 1: class1\\
\noindent Image 2: class2\\
\noindent Please first output one line for the label of image 3. In the subsequent line, please provide a comprehensive explanation of your classification." 
\end{mdframed}

\begin{mdframed}
\noindent\textbf{In-context learning 3 (ICL3).}\\
\noindent Combining three images on one figure, two for in-context learning and one for predicting. Upload 3 combined figures (batch size = 3) on GPT4V.\\
\noindent \textbf{Prompt:} ``Instruction: : classify the images into two classes for each group {class1, class2}, generate 4 results. 

\noindent Example: the label of the above images:\\
\noindent Image 1: class1\\
\noindent Image 2: class2\\
\noindent Please first output one line for the label of image 3. In the subsequent line, please provide a comprehensive explanation of your classification." 
\end{mdframed}



\begin{mdframed}
\noindent\textbf{In-context learning 4 (ICL4).}\\
\noindent Combining 9 images in one figure and upload it on GPT4V, 6 of them are for in-context learning and the rest 3 for predicting.\\
\noindent \textbf{Prompt:} ``Instruction: classify the images into two classes {class1, class2}

\noindent Example: the label of the above images:\\
\noindent Image 1: class1\\
\noindent Image 2: class1\\
\noindent Image 3: class1\\
\noindent Image 4: class2\\
\noindent Image 5: class2\\
\noindent Image 6: class2\\
\noindent Please first output one line for the label of image 7, image 8 and image 9. In the subsequent line, please provide a comprehensive explanation of your classification." 
\end{mdframed}

\begin{mdframed}
\noindent\textbf{In-context learning with reasoning 1 (ICL-R1).}\\
\noindent Upload three images separately on GPT4V, two for in-context learning and one for predicting.\\
\noindent \textbf{Prompt:} ``Instruction: classify the images into two classes {class1, class2}

\noindent Example: the label of the above images:\\
\noindent Image 1: class1\\
\noindent Image 2: class2\\
\noindent Explanation: In image 1 we can observe ..., but in image 2 we don't have such observation. Thus we classified image 1 as class1 and image 2 as class2. \\
\noindent Please provide the classification of Image 3 in one line, taking into account the observed patterns in Image 3. Following that, offer a detailed explanation step-by-step.
\end{mdframed}

\begin{mdframed}
\noindent\textbf{In-context learning with reasoning 2 (ICL-R2).}\\
\noindent Combining 9 images in one figure and upload it on GPT4V, 6 of them are for in-context learning and the rest 3 for predicting.\\
\noindent \textbf{Prompt:} ``Instruction: classify the images into two classes {class1, class2}

\noindent Example: the label of the above images:\\
\noindent Image 1: class1\\
\noindent Image 2: class1\\
\noindent Image 3: class1\\
\noindent Image 4: class2\\
\noindent Image 5: class2\\
\noindent Image 6: class2\\
\noindent Explanation: In image 1-3 we can observe ... but in image 2 we don't have such observation. \\
\noindent Please first output one line for the label of image 7, image 8 and image 9. In the subsequent line, please provide a comprehensive explanation of your classification." 
\end{mdframed}







The crux of our method lies in this nuanced interaction with the model. Our results elucidate that the strategic construction of prompts—capitalizing on the model's inherent language and reasoning capabilities—enables GPT-4V to perform on par with established medical image classification benchmarks. This finding not only underscores the versatility of large language models but also heralds a potentially transformative application in the medical imaging domain.

\section{Experiment}
In the experiment, we test our proposed prompt on the open sourced Kaggle COVID-19 lung X-ray dataset. This dataset contains 181 training examples, 111 of them are COVID and the rest are normal case. There are total 46 examples in the test set, where 26 of them are COVID case and the rest are normal case.

\noindent{\textbf{Baseline Settings.}} We construct baselines using Convolution Neural Network based backbones to demonstrate the effectiveness of our method in the few-shot learning setting. Following previous works~\cite{liu2021magnetic, odusami2021analysis}, we use ResNet-18(RN-18)~\cite{he2016deep} and VGG-16~\cite{simonyan2014very} pre-trained on ImageNet-1k~\cite{imagenet15russakovsky} as our image classifier. Output dimension of the final fully-connected layer is set to 2 to fit the binary classification task. During training, we optimize the model using SGD for 20 epoches with batch size of 2, and decrease the learning rate by 5 after epoch 10 and 15 respectively. For better convergence, initial learning rate is set to 0.1 for ResNet-18 and 0.001 for VGG-16. We also apply a simple augmentation technique of random rotation and center cropping to improve model robustness. For the few-shot setting, we randomly select 6 images (3 covid, 3 normal) for training. We repeat the experiments for 5 times with different random seed and report the average result.

\begin{table}[!htbp]
\caption{Main result}
\label{table}
\resizebox{1\textwidth}{!}{
\begin{tabular}{c|ccc|ccc|c}
\toprule
       & \multicolumn{3}{c|}{COVID}                                          & \multicolumn{3}{c|}{Normal}                                         & All      \\ \midrule
       & \multicolumn{1}{c|}{Precision} & \multicolumn{1}{c|}{Recall} & F1   & \multicolumn{1}{c|}{Precision} & \multicolumn{1}{c|}{Recall} & F1   & Accuracy \\ \midrule
Naive  & \multicolumn{1}{c|}{0.83}      & \multicolumn{1}{c|}{0.77}   & 0.80 & \multicolumn{1}{c|}{0.73}      & \multicolumn{1}{c|}{0.80}   & 0.76 & 0.78     \\ 
ICL1   & \multicolumn{1}{c|}{0.76}      & \multicolumn{1}{c|}{0.73}   & 0.75 & \multicolumn{1}{c|}{0.67}      & \multicolumn{1}{c|}{0.70}   & 0.68 & 0.72     \\ 
ICL2   & \multicolumn{1}{c|}{1.00}      & \multicolumn{1}{c|}{0.62}   & 0.76 & \multicolumn{1}{c|}{0.66}      & \multicolumn{1}{c|}{1.00}   & 0.80 & 0.78     \\ 
ICL3   & \multicolumn{1}{c|}{0.90}      & \multicolumn{1}{c|}{0.69}   & 0.78 & \multicolumn{1}{c|}{0.69}      & \multicolumn{1}{c|}{0.90}   & 0.78 & 0.83     \\ 
ICL4   & \multicolumn{1}{c|}{0.95}      & \multicolumn{1}{c|}{0.77}   & 0.85 & \multicolumn{1}{c|}{0.76}      & \multicolumn{1}{c|}{0.95}   & 0.84 & 0.85     \\ 
ICL-R1 & \multicolumn{1}{c|}{0.67}      & \multicolumn{1}{c|}{0.80}   & 0.73 & \multicolumn{1}{c|}{0.82}      & \multicolumn{1}{c|}{0.69}   & 0.75 & 0.74     \\ 
ICL-R2 & \multicolumn{1}{c|}{0.83}      & \multicolumn{1}{c|}{0.75}   & 0.79 & \multicolumn{1}{c|}{0.82}      & \multicolumn{1}{c|}{0.88}   & 0.85 & 0.83     \\ \midrule
RN-18\cite{he2016deep} (Full)  & \multicolumn{1}{c|}{0.92}          & \multicolumn{1}{c|}{0.96}      &    0.94  & \multicolumn{1}{c|}{0.95}          & \multicolumn{1}{c|}{0.90}       &    0.92  &   0.93   \\ 

RN-18\cite{he2016deep} (6-shot)  & \multicolumn{1}{c|}{0.85}          & \multicolumn{1}{c|}{0.65}       &   0.74   & \multicolumn{1}{c|}{0.65}          & \multicolumn{1}{c|}{0.85}       &    0.74  &  0.74   \\ 
VGG16\cite{simonyan2014very} (Full)  & \multicolumn{1}{c|}{1.00}          & \multicolumn{1}{c|}{0.92}      &    0.96  & \multicolumn{1}{c|}{0.91}          & \multicolumn{1}{c|}{1.00}       &    0.95  &   0.96   \\ 

VGG16\cite{simonyan2014very} (6-shot)  & \multicolumn{1}{c|}{1.0}          & \multicolumn{1}{c|}{0.65}       &   0.79   & \multicolumn{1}{c|}{0.69}          & \multicolumn{1}{c|}{1.0}       &   0.81   &   0.80  \\ \bottomrule
\end{tabular}
}
\end{table}

\noindent{\textbf{Result analysis.}}
Consolidating all images into a single figure has demonstrated enhanced performance compared to uploading them individually. This improvement could potentially be attributed to the focused attention mechanism of GPT-4V, which, when presented with separate images, might concentrate disproportionately on specific images, consequently leading to biased outcomes.

GPT-4V exhibits superior performance compared to the few-shot baseline when provided with an equivalent number of training instances. However, it does not yet match the efficacy of the comprehensive baseline model that benefits from training on the complete set of examples. This indicates that while GPT-4V's adaptability is promising, certain optimizations might be necessary to fully realize its learning potential.

Contrary to expectations, supplementing the GPT-4V prompts with reasons underlying the classifications does not yield an improvement in results. This may be due to a misalignment between the provided reasoning and the model's processing capabilities, suggesting that the reasons integrated into the prompts were either not appropriately formulated for GPT-4V's comprehension or that the model currently lacks the capacity to incorporate such reasoning effectively into its decision-making process. Further investigations are needed to uncover the intricacies of this observation.

\section{Conclusion}
In conclusion, we take the first step into the application of GPT-4V for medical image classification. By employing in-context learning, this study circumvented traditional limitations associated with deep learning models, particularly the necessity for extensive, task-specific training and vast labeled datasets. The tailored prompts guided GPT-4V to effectively interpret and analyze medical images, achieving a level of accuracy on par with conventional methods. This finding underscores the versatile potential of large foundation models in medical diagnostics and opens the door to further innovations that could reshape the landscape of healthcare, making it more intuitive, accessible, and reliable. Beyond just technical implications, the success of this approach advocates for a future where AI's role extends from being a mere tool to an adaptable ally, capable of navigating the nuanced and critical terrains of patient care.

{\small
\bibliographystyle{unsrt}
\bibliography{nips}

\begin{thebibliography}{10}

\bibitem{simonyan2014very}
Karen Simonyan and Andrew Zisserman.
\newblock Very deep convolutional networks for large-scale image recognition.
\newblock {\em arXiv preprint arXiv:1409.1556}, 2014.

\bibitem{szegedy2015going}
Christian Szegedy, Wei Liu, Yangqing Jia, Pierre Sermanet, Scott Reed, Dragomir Anguelov, Dumitru Erhan, Vincent Vanhoucke, and Andrew Rabinovich.
\newblock Going deeper with convolutions.
\newblock In {\em Proceedings of the IEEE conference on computer vision and pattern recognition}, pages 1--9, 2015.

\bibitem{he2016deep}
Kaiming He, Xiangyu Zhang, Shaoqing Ren, and Jian Sun.
\newblock Deep residual learning for image recognition.
\newblock In {\em Proceedings of the IEEE conference on computer vision and pattern recognition}, pages 770--778, 2016.

\bibitem{huang2017densely}
Gao Huang, Zhuang Liu, Laurens Van Der~Maaten, and Kilian~Q Weinberger.
\newblock Densely connected convolutional networks.
\newblock In {\em Proceedings of the IEEE conference on computer vision and pattern recognition}, pages 4700--4708, 2017.

\bibitem{cai2020review}
Lei Cai, Jingyang Gao, and Di~Zhao.
\newblock A review of the application of deep learning in medical image classification and segmentation.
\newblock {\em Annals of translational medicine}, 8(11), 2020.

\bibitem{kim2022transfer}
Hee~E Kim, Alejandro Cosa-Linan, Nandhini Santhanam, Mahboubeh Jannesari, Mate~E Maros, and Thomas Ganslandt.
\newblock Transfer learning for medical image classification: a literature review.
\newblock {\em BMC medical imaging}, 22(1):69, 2022.

\bibitem{huang2023self}
Shih-Cheng Huang, Anuj Pareek, Malte Jensen, Matthew~P Lungren, Serena Yeung, and Akshay~S Chaudhari.
\newblock Self-supervised learning for medical image classification: a systematic review and implementation guidelines.
\newblock {\em NPJ Digital Medicine}, 6(1):74, 2023.

\bibitem{nayem2023few}
Jannatul Nayem, Sayed~Sahriar Hasan, Noshin Amina, Bristy Das, Md~Shahin Ali, Md~Manjurul Ahsan, and Shivakumar Raman.
\newblock Few shot learning for medical imaging: A comparative analysis of methodologies and formal mathematical framework.
\newblock {\em arXiv preprint arXiv:2305.04401}, 2023.

\bibitem{liu-etal-2022-makes}
Jiachang Liu, Dinghan Shen, Yizhe Zhang, Bill Dolan, Lawrence Carin, and Weizhu Chen.
\newblock What makes good in-context examples for {GPT}-3?
\newblock In {\em Proceedings of Deep Learning Inside Out (DeeLIO 2022): The 3rd Workshop on Knowledge Extraction and Integration for Deep Learning Architectures}, pages 100--114, Dublin, Ireland and Online, May 2022. Association for Computational Linguistics.

\bibitem{wei2021finetuned}
Jason Wei, Maarten Bosma, Vincent~Y. Zhao, Kelvin Guu, Adams~Wei Yu, Brian Lester, Nan Du, Andrew~M. Dai, and Quoc~V. Le.
\newblock Finetuned language models are zero-shot learners, 2021.

\bibitem{wang2023large}
Xinyi Wang, Wanrong Zhu, Michael Saxon, Mark Steyvers, and William~Yang Wang.
\newblock Large language models are latent variable models: Explaining and finding good demonstrations for in-context learning, 2023.

\bibitem{chen2023demonstrations}
Jiuhai Chen, Lichang Chen, Chen Zhu, and Tianyi Zhou.
\newblock How many demonstrations do you need for in-context learning?, 2023.

\bibitem{role2022}
Sewon Min, Xinxi Lyu, Ari Holtzman, Mikel Artetxe, Mike Lewis, Hannaneh Hajishirzi, and Luke Zettlemoyer.
\newblock Rethinking the role of demonstrations: What makes in-context learning work?
\newblock {\em Proceedings of the 2022 Conference on Empirical Methods in Natural Language Processing}, 2022.

\bibitem{brown2020language}
Tom Brown, Benjamin Mann, Nick Ryder, Melanie Subbiah, Jared~D Kaplan, Prafulla Dhariwal, Arvind Neelakantan, Pranav Shyam, Girish Sastry, Amanda Askell, et~al.
\newblock Language models are few-shot learners.
\newblock {\em Advances in neural information processing systems}, 33:1877--1901, 2020.

\bibitem{chowdhery2022palm}
Aakanksha Chowdhery, Sharan Narang, Jacob Devlin, Maarten Bosma, Gaurav Mishra, Adam Roberts, Paul Barham, Hyung~Won Chung, Charles Sutton, Sebastian Gehrmann, Parker Schuh, Kensen Shi, Sasha Tsvyashchenko, Joshua Maynez, Abhishek Rao, Parker Barnes, Yi~Tay, Noam Shazeer, Vinodkumar Prabhakaran, Emily Reif, Nan Du, Ben Hutchinson, Reiner Pope, James Bradbury, Jacob Austin, Michael Isard, Guy Gur{-}Ari, Pengcheng Yin, Toju Duke, Anselm Levskaya, Sanjay Ghemawat, Sunipa Dev, Henryk Michalewski, Xavier Garcia, Vedant Misra, Kevin Robinson, Liam Fedus, Denny Zhou, Daphne Ippolito, David Luan, Hyeontaek Lim, Barret Zoph, Alexander Spiridonov, Ryan Sepassi, David Dohan, Shivani Agrawal, Mark Omernick, Andrew~M. Dai, Thanumalayan~Sankaranarayana Pillai, Marie Pellat, Aitor Lewkowycz, Erica Moreira, Rewon Child, Oleksandr Polozov, Katherine Lee, Zongwei Zhou, Xuezhi Wang, Brennan Saeta, Mark Diaz, Orhan Firat, Michele Catasta, Jason Wei, Kathy Meier{-}Hellstern, Douglas Eck, Jeff Dean, Slav Petrov, and Noah Fiedel.
\newblock Palm: Scaling language modeling with pathways.
\newblock {\em CoRR}, abs/2204.02311, 2022.

\bibitem{thoppilan2022lamda}
Romal Thoppilan, Daniel~De Freitas, Jamie Hall, Noam Shazeer, Apoorv Kulshreshtha, Heng{-}Tze Cheng, Alicia Jin, Taylor Bos, Leslie Baker, Yu~Du, YaGuang Li, Hongrae Lee, Huaixiu~Steven Zheng, Amin Ghafouri, Marcelo Menegali, Yanping Huang, Maxim Krikun, Dmitry Lepikhin, James Qin, Dehao Chen, Yuanzhong Xu, Zhifeng Chen, Adam Roberts, Maarten Bosma, Yanqi Zhou, Chung{-}Ching Chang, Igor Krivokon, Will Rusch, Marc Pickett, Kathleen~S. Meier{-}Hellstern, Meredith~Ringel Morris, Tulsee Doshi, Renelito~Delos Santos, Toju Duke, Johnny Soraker, Ben Zevenbergen, Vinodkumar Prabhakaran, Mark Diaz, Ben Hutchinson, Kristen Olson, Alejandra Molina, Erin Hoffman{-}John, Josh Lee, Lora Aroyo, Ravi Rajakumar, Alena Butryna, Matthew Lamm, Viktoriya Kuzmina, Joe Fenton, Aaron Cohen, Rachel Bernstein, Ray Kurzweil, Blaise Aguera{-}Arcas, Claire Cui, Marian Croak, Ed~H. Chi, and Quoc Le.
\newblock Lamda: Language models for dialog applications.
\newblock {\em CoRR}, abs/2201.08239, 2022.

\bibitem{rae2021scaling}
Jack~W. Rae, Sebastian Borgeaud, Trevor Cai, Katie Millican, Jordan Hoffmann, H.~Francis Song, John Aslanides, Sarah Henderson, Roman Ring, Susannah Young, Eliza Rutherford, Tom Hennigan, Jacob Menick, Albin Cassirer, Richard Powell, George van~den Driessche, Lisa~Anne Hendricks, Maribeth Rauh, Po{-}Sen Huang, Amelia Glaese, Johannes Welbl, Sumanth Dathathri, Saffron Huang, Jonathan Uesato, John Mellor, Irina Higgins, Antonia Creswell, Nat McAleese, Amy Wu, Erich Elsen, Siddhant~M. Jayakumar, Elena Buchatskaya, David Budden, Esme Sutherland, Karen Simonyan, Michela Paganini, Laurent Sifre, Lena Martens, Xiang~Lorraine Li, Adhiguna Kuncoro, Aida Nematzadeh, Elena Gribovskaya, Domenic Donato, Angeliki Lazaridou, Arthur Mensch, Jean{-}Baptiste Lespiau, Maria Tsimpoukelli, Nikolai Grigorev, Doug Fritz, Thibault Sottiaux, Mantas Pajarskas, Toby Pohlen, Zhitao Gong, Daniel Toyama, Cyprien de~Masson~d'Autume, Yujia Li, Tayfun Terzi, Vladimir Mikulik, Igor Babuschkin, Aidan Clark, Diego de~Las~Casas, Aurelia Guy,
  Chris Jones, James Bradbury, Matthew~J. Johnson, Blake~A. Hechtman, Laura Weidinger, Iason Gabriel, William~S. Isaac, Edward Lockhart, Simon Osindero, Laura Rimell, Chris Dyer, Oriol Vinyals, Kareem Ayoub, Jeff Stanway, Lorrayne Bennett, Demis Hassabis, Koray Kavukcuoglu, and Geoffrey Irving.
\newblock Scaling language models: Methods, analysis {\&} insights from training gopher.
\newblock {\em CoRR}, abs/2112.11446, 2021.

\bibitem{radford2021learning}
Alec Radford, Jong~Wook Kim, Chris Hallacy, Aditya Ramesh, Gabriel Goh, Sandhini Agarwal, Girish Sastry, Amanda Askell, Pamela Mishkin, Jack Clark, et~al.
\newblock Learning transferable visual models from natural language supervision.
\newblock In {\em International conference on machine learning}, pages 8748--8763. PMLR, 2021.

\bibitem{finn2017model}
Chelsea Finn, Pieter Abbeel, and Sergey Levine.
\newblock Model-agnostic meta-learning for fast adaptation of deep networks.
\newblock In {\em International conference on machine learning}, pages 1126--1135. PMLR, 2017.

\bibitem{dalal2005histograms}
Navneet Dalal and Bill Triggs.
\newblock Histograms of oriented gradients for human detection.
\newblock In {\em 2005 IEEE computer society conference on computer vision and pattern recognition (CVPR'05)}, volume~1, pages 886--893. Ieee, 2005.

\bibitem{he1990texture}
Dong-Chen He and Li~Wang.
\newblock Texture unit, texture spectrum, and texture analysis.
\newblock {\em IEEE transactions on Geoscience and Remote Sensing}, 28(4):509--512, 1990.

\bibitem{ganjdanesh2022multi}
Alireza Ganjdanesh, Jipeng Zhang, Wei Chen, and Heng Huang.
\newblock Multi-modal genotype and phenotype mutual learning to enhance single-modal input based longitudinal outcome prediction.
\newblock In {\em International Conference on Research in Computational Molecular Biology}, pages 209--229. Springer, 2022.

\bibitem{TAPE}
Yanshuo Chen, Yixuan Wang, Yuelong Chen, Yuqi Cheng, Yumeng Wei, Yunxiang Li, Jiuming Wang, Yingying Wei, Ting-Fung Chan, and Yu~Li.
\newblock Deep autoencoder for interpretable tissue-adaptive deconvolution and cell-type-specific gene analysis.
\newblock {\em Nature Communications}, 13(1):6735, 2022.

\bibitem{kong2022classification}
Lingzhi Kong and Jinyong Cheng.
\newblock Classification and detection of covid-19 x-ray images based on densenet and vgg16 feature fusion.
\newblock {\em Biomedical Signal Processing and Control}, 77:103772, 2022.

\bibitem{liu2021magnetic}
Yan Liu, Guo-rong She, and Shu-xaing Chen.
\newblock Magnetic resonance image diagnosis of femoral head necrosis based on resnet18 network.
\newblock {\em Computer Methods and Programs in Biomedicine}, 208:106254, 2021.

\bibitem{odusami2021analysis}
Modupe Odusami, Rytis Maskeli{\=u}nas, Robertas Dama{\v{s}}evi{\v{c}}ius, and Tomas Krilavi{\v{c}}ius.
\newblock Analysis of features of alzheimer’s disease: Detection of early stage from functional brain changes in magnetic resonance images using a finetuned resnet18 network.
\newblock {\em Diagnostics}, 11(6):1071, 2021.

\bibitem{imagenet15russakovsky}
Olga Russakovsky, Jia Deng, Hao Su, Jonathan Krause, Sanjeev Satheesh, Sean Ma, Zhiheng Huang, Andrej Karpathy, Aditya Khosla, Michael Bernstein, Alexander~C. Berg, and Li~Fei-Fei.
\newblock {ImageNet Large Scale Visual Recognition Challenge}.
\newblock {\em International Journal of Computer Vision (IJCV)}, 115(3):211--252, 2015.

\end{thebibliography}
}

\end{document}